\documentclass[preprint,aps,prd,superscriptaddress,groupedaddress,showpacs]{revtex4-1}
\usepackage{amsmath}
\usepackage{graphicx}
\usepackage{dcolumn}
\usepackage{bm}
\usepackage{amssymb}

\begin{document}

\title{Origin of the Relaxation Time in Dissipative Fluid Dynamics}

\author{Gabriel S.\ Denicol}

\affiliation{Institut f\"ur Theoretische Physik, Goethe University, 60438
Frankfurt am Main, Germany}
\author{Jorge Noronha}

\affiliation{Department of Physics, Columbia University, 
New York, NY 10027, USA, and 
\\
Instituto de F\'{\i}sica, Universidade Federal do Rio de Janeiro, C.
P. 68528, 21945-970, Rio de Janeiro, Brasil}
\author{Harri Niemi}

\affiliation{Frankfurt Institute for Advanced Studies (FIAS),
60438 Frankfurt am Main, Germany}
\author{Dirk H.\ Rischke}

\affiliation{Institut f\"ur Theoretische Physik, Goethe University, and
\\
Frankfurt Institute for Advanced Studies (FIAS), 60438 Frankfurt am Main, Germany }

\begin{abstract}
We show how the linearized equations of motion of any dissipative current
are determined by the analytical structure of the associated retarded
Green's function. If the singularity of the Green's function, which is
nearest to the origin in the complex-frequency plane, is a simple pole on
the imaginary frequency axis, the linearized equations of motion can be
reduced to relaxation-type equations for the dissipative currents. The value
of the relaxation time is given by the inverse of this pole. We prove that,
if the relaxation time is sent to zero, or equivalently, the pole to
infinity, the dissipative currents approach the values given by the standard
gradient expansion.
\end{abstract}

\maketitle


\section{Introduction}

During the last decade, relativistic fluid dynamics has been used as one of
the main theoretical tools \cite{Heinz:2009xj} in the study of the hot and
dense matter created in heavy-ion collisions at RHIC and, most recently, at
LHC. While the properties of non-relativistic fluids are now well understood 
\cite{landau, Torr}, the non-trivial physical consequences imposed by
causality and stability of relativistic fluids are still under intense
theoretical investigation \cite{hiscock,us}.

In non-relativistic Navier-Stokes theory, the dissipative currents, such as
the bulk viscous pressure, the shear stress tensor, and the heat flow, are
assumed to be linearly proportional to the fluid-dynamical forces, such as
spatial gradients of fluid velocity, temperature, and chemical potential.
The constants of proportionality are the bulk viscosity, the shear
viscosity, and the heat conductivity. Navier-Stokes theory can be extended
by considering higher-order gradients of fluid velocity, temperature, or
chemical potential, leading to the Burnett equations (including second-order
gradients), the super-Burnett equations (including third-order gradients)
etc. \cite{Burnett}. A systematic derivation of these equations is provided
by the gradient expansion. The first-order truncation of the gradient
expansion, i.e., Navier-Stokes theory, is stable, but not causal, as it
allows for propagation of signals with infinite speed \cite{hiscock}.
Higher-order truncations suffer from the Bobylev instability \cite{Bobylev},
and are thus neither stable nor causal.

Eckart, and later, Landau and Lifshitz were the first to attempt a
relativistic formulation of fluid dynamics \cite{eckart,landau}. Their
derivation was based on a relativistically covariant extension of
traditional Navier-Stokes theory. The resulting equations coincide with a
first-order truncation of a relativistic formulation of the gradient
expansion. Unlike the non-relativistic case, however, already the
first-order truncation of the gradient expansion, i.e., the relativistic
generalization of Navier-Stokes theory, is not only acausal but,
unfortunately, also unstable \cite{hiscock}.

Israel and Stewart \cite{IS} were among the first to formulate a
relativistic theory of fluid dynamics that respected causality and was
potentially stable. Their formulation was a relativistic extension of Grad's
theory \cite{Grad}, where the fluid-dynamical dissipative currents appear as
dynamical variables which relax to the values of Navier-Stokes theory on
characteristic time scales, usually referred to as relaxation times. Thus,
unlike for the gradient expansion the dissipative currents in Israel-Stewart
theory (IS) do not have to be zero in the absence of gradients. Instead,
they decay to zero on the time scales given by the relaxation times.

One of the features of Israel-Stewart theory is that it reduces to
Navier-Stokes theory in the limit of vanishing relaxation times. In other
words, in Navier-Stokes theory the dissipative currents relax
instantaneously to the values given by the fluid-dynamical forces, which
leads to a violation of causality. For the theory to be causal it is
therefore \textit{necessary\/} that the relaxation times assume a non-zero
value, but this is \textit{not sufficient}. It was shown in Ref.\ \cite{us}
for the case of bulk and shear viscosity that causality imposes a \textit{%
stronger\/} constraint for IS theories: the ratio of the relaxation times to
the viscosity coefficients must \textit{exceed certain values\/} \cite{us}.
For instance for the case of shear viscosity only, the ratio of shear
relaxation time $\tau _{\pi }$ to shear viscosity $\eta $ must obey the
relation 
\begin{equation}
\frac{\tau _{\pi }}{\eta /(\varepsilon +P)}\geq \frac{4}{3(1-c_{s}^{2})}\;,
\label{asc}
\end{equation}%
where $\varepsilon $, $P$, and $c_{s}$ are the energy density, the
thermodynamic pressure, and the velocity of sound, respectively. It was also
shown in Ref.\ \cite{us} that, for relativistic fluids, the causality
condition (\ref{asc}) also implies stability of the fluid-dynamical
equations. Any \textit{physically meaningful\/} theory of fluid dynamics
must not be intrinsically unstable, and any relativistic generalization of
such a physically meaningful theory must be causal. Therefore, Eq.\ (\ref%
{asc}) implies that, for any non-zero value of viscosity, one \textit{%
cannot\/} take the limit of vanishing relaxation time. In this sense, the
transient dynamics of the dissipative currents cannot be neglected when
dealing with relativistic fluids. This realization forms the basis for the
following discussion.

In this paper, we derive the equations of motion for dissipative currents in
the linear regime. We show how the linearized equation of motion of any
dissipative current is determined once the analytical structure of the
associated retarded Green's function is known. In the case that the
singularity of the retarded Green's function nearest to the origin is a
simple pole on the imaginary axis, we prove that equation of motion can be
reduced to a relaxation-type equation. The relaxation time is equal to minus
the inverse of the imaginary part of this pole. We prove that this
prescription gives a value for the relaxation time that, under certain
simplifying assumptions, coincides with the one derived by matching
relativistic fluid dynamics to kinetic theory. We also demonstrate that
attempts to derive the relaxation time from considering the long-wavelength,
low-frequency (i.e., fluid-dynamical) limit of the retarded Green's function
in general fail to give the correct result. This shows that transient
dynamics is determined by the slowest microscopic and not by the fastest
fluid-dynamical time scale.

This paper is organized as follows. In the next section we define the
notation used throughout this paper. In Sec.\ \ref{gradientsection} we show
the equivalency of the gradient expansion in coordinate space with a Taylor
series in momentum space. In Sec.\ \ref{polessection} we derive the main
results of this paper and establish the connection between the analytical
structure of retarded Green's functions and the linear transport
coefficients in fluid dynamics. We show in Secs.\ \ref{Boltzmannsection} and %
\ref{metricsection} that the linear transport coefficients computed using
either the linearized Boltzmann equation or via the disturbances in the
space-time metric follow the general method and formulae derived in Sec.\ %
\ref{polessection}. In Sec.\ \ref{Discussion} we compare our results to
previous derivations of relaxation-type equations for the dissipative
currents based on the gradient expansion. We conclude this paper with a
summary of our results.

\section{Definitions}

Let us consider a general linear relation between a dissipative current $%
J\left(X\right)$ and a thermodynamical force $F\left(X\right) $,%
\begin{equation}
J\left( X\right) =\int d^{4}X^{\prime }\,G_{R}\left( X- X^{\prime}\right) \,
F\left( X^{\prime }\right) \text{,}  \label{LinearRel}
\end{equation}%
where $X=(t,\mathbf{x})$ is the coordinate four-vector in space-time. Any
translationally invariant theory that has a linear relation between $J$ and $%
F$ can always be written in the form of Eq.\ (\ref{LinearRel}). In fluid
dynamics, $J$ could e.g.\ be the shear stress tensor $\pi ^{\mu \nu}$ and $F$
the shear tensor $\sigma ^{\mu \nu }$, or $J$ could be the diffusion current
of a density $n$ and $F\sim \partial n$.

We define our Fourier transformation in the following way 
\begin{eqnarray}
\tilde{A}\left( Q\right) &=&\int d^{4}X\exp \left( iQ\cdot X\right) A\left(
X\right) \;, \\
A\left( X\right) &=&\int \frac{d^{4}Q}{\left( 2\pi \right) ^{4}}\exp \left(
-iQ\cdot X\right) \tilde{A}\left( Q\right) \;.  \label{4}
\end{eqnarray}%
Here, $Q=\left( \omega ,\mathbf{q}\right) $ is the momentum four-vector, and 
$Q\cdot X\equiv q^{\mu }x_{\mu }$ is the scalar four-product of the
four-momentum vector $Q$ with the coordinate four-vector $X$. Our metric
signature is $\eta _{\mu \nu }=\mathrm{diag}\,(+,-,-,-)$. Using this
convention, we can rewrite Eq.\ (\ref{LinearRel}) in terms of the Fourier
transforms of the retarded Green's function and of the thermodynamic force, $%
\tilde{G}_{R}\left( Q\right) $ and $\tilde{F}\left( Q\right) $,
respectively, which then gives 
\begin{equation}
\tilde{J}\left( Q\right) =\tilde{G}_{R}\left( Q\right) \tilde{F}\left(
Q\right) \text{ \ }.  \label{LinRelFou}
\end{equation}%
We only consider systems where the microscopic dynamics is invariant under
time reversal and, thus, $\mathrm{Re}\,\tilde{G}_{R}$ is an even function of 
$\omega $, while $\mathrm{Im}\,\tilde{G}_{R}$ is an odd function of $\omega $%
. Note that Eq.\ (\ref{LinRelFou}) implies that the current $J$ can also be
expressed as an integral over $Q$,
\begin{equation}
J\left( X\right) =\int \frac{d^{4}Q}{(2\pi )^{4}}\,\exp \left( -iQ\cdot
X\right) \tilde{G}_{R}\left( Q\right) \tilde{F}\left( Q\right) \;.  \label{6}
\end{equation}%
Thus, $\tilde{G}_{R}$ at \textit{all} frequencies can contribute to the
dynamics of $J$. Equations (\ref{LinearRel}), (\ref{LinRelFou}), and (\ref{6}%
) are, of course, equivalent and contain all the information about the
underlying microscopic theory that can be obtained through a linear
analysis. In this work, we show how the analytical structure of the retarded
Green's function $\tilde{G}_{R}$ determines the equation of motion for the
current $J$.

\section{Equivalency between gradient expansion and Taylor series}

\label{gradientsection}

In this section, we show that the gradient expansion in space-time is
actually equivalent to a Taylor series in 4-momentum space. For the sake of
simplicity, we suppress any dependence on spatial coordinates or,
equivalently, on 3-momentum, retaining only the dependence on time $t$ and
(complex) frequency $\omega$. The coordinate, or 3-momentum dependence,
respectively, will be restored later. We implicitly work in the rest frame
of the fluid, such that the equations of motion do not appear to be
relativistically covariant, however, covariance can be restored by a proper
Lorentz-boost.

Let us assume that $\tilde{G}_{R}\left( \omega \right) $ is analytic in the
whole complex $\omega $ plane. This can be considered as the limiting case
when $\tilde{G}_{R}\left( \omega \right) $ has singularities, but all of
them are pushed to infinity by some suitable limiting procedure. In this
situation a Taylor expansion of $\tilde{G}_{R}\left( \omega \right) $ around
the origin has infinite convergence radius and thus provides a valid
representation of $\tilde{G}_{R}\left( \omega \right) $ in the whole complex
plane, 
\begin{equation}
\tilde{G}_{R}\left( \omega \right) =\tilde{G}_{R}\left( 0\right) +\partial
_{\omega }\left. \tilde{G}_{R}\left( \omega \right) \right\vert _{\omega
=0}\omega +\frac{1}{2}\partial _{\omega }^{2}\left. \tilde{G}_{R}\left(
\omega \right) \right\vert _{\omega =0}\omega ^{2}+\mathcal{O}\left( \omega
^{3}\right) \,.  \label{Taylor1}
\end{equation}%
Using Eqs.\ (\ref{4}) and (\ref{Taylor1}), as well as the Fourier
representation of Dirac's delta function 
\begin{equation}
\delta \left( t-t^{\prime }\right) =\int \frac{d\omega }{2\pi }\,\exp \left[
-i\omega \left( t-t^{\prime }\right) \right] ,  \label{Dirac}
\end{equation}%
it is straightforward to obtain the general form of $G_{R}\left( t-t^{\prime
}\right) $, 
\begin{eqnarray}
G_{R}\left( t-t^{\prime }\right) &=&\tilde{G}_{R}\left( 0\right) \delta
\left( t-t^{\prime }\right) +i\partial _{\omega }\left. \tilde{G}_{R}\left(
\omega \right) \right\vert _{\omega =0}\partial _{t}\delta \left(
t-t^{\prime }\right)  \notag \\
&&-\frac{1}{2}\partial _{\omega }^{2}\left. \tilde{G}_{R}\left( \omega
\right) \right\vert _{\omega =0}\partial _{t}^{2}\delta \left( t-t^{\prime
}\right) +\mathcal{O}\left( \partial _{t}^{3}\right) \;.  \label{GRt}
\end{eqnarray}%
Substituting Eq.\ (\ref{GRt}) into Eq.\ (\ref{LinearRel}) we obtain the
following equation of motion for the dissipative current, 
\begin{equation}
J\left( t\right) =\bar{D}_{0}F\left( t\right) +\bar{D}_{1}\partial
_{t}F\left( t\right) +\bar{D}_{2}\partial _{t}^{2}F\left( t\right) +\mathcal{%
O}\left( \partial _{t}^{3}F\right) ,\,  \label{ExpJ}
\end{equation}%
where we introduced the coefficients%
\begin{eqnarray}
\bar{D}_{0} &=&\tilde{G}_{R}\left( 0\right) ,\text{ }  \notag \\
\bar{D}_{1} &=&i\partial _{\omega }\left. \tilde{G}_{R}\left( \omega \right)
\right\vert _{\omega =0},  \label{Coeffs} \\
\bar{D}_{2} &=&-\frac{1}{2}\partial _{\omega }^{2}\left. \tilde{G}_{R}\left(
\omega \right) \right\vert _{\omega =0}\;,  \notag \\
& \vdots &
\end{eqnarray}%
This is nothing but the so-called gradient expansion, in which the current $%
J $ is expressed in terms of the thermodynamic force $F$ and its
derivatives. Note that the standard gradient expansion does not involve time
derivatives of the thermodynamic force. This is not a problem, since one can
always replace time derivatives with spatial gradients using the
conservation equations of fluid dynamics.

It is important to remark that when the system exhibits a clear separation
between the typical microscopic and macroscopic scales, $\lambda $ and $\ell 
$, respectively, it is possible to truncate the expansion on the right-hand
side of Eq.\ (\ref{ExpJ}). A microscopic scale is, for example, the
mean-free path in dilute gases. A macroscopic scale is given by the inverse
of the gradient of a macroscopic variable, such as energy density, charge
density, or fluid velocity. Note that the thermodynamic force, $F$, is
already proportional to a gradient of a macroscopic variable, and thus $%
F\sim \ell ^{-1}$. Every additional derivative $\partial _{t}$ brings in
another inverse power of $\ell $, $\partial _{t}^{n}F\sim \ell ^{-(n+1)}$.
The microscopic scale $\lambda $ is contained in $\tilde{G}_{R}$ and its
derivatives with respect to $\omega $. Thus, up to some overall power of $%
\lambda $ (which restores the correct scaling dimension), $\tilde{G}%
_{R}(0)\sim \lambda $, and each additional derivative $\partial _{\omega }$
brings in another power of $\lambda $, such that $\bar{D}_{n}\sim \lambda
^{n+1}$. Therefore, the terms $\bar{D}_{0}F$, $\bar{D}_{1}\partial _{t}F$,
and $\bar{D}_{2}\partial _{t}^{2}F$ in Eq.\ (\ref{ExpJ}) are of order $%
\lambda /\ell $, $\left( \lambda /\ell \right) ^{2}$ and $\left( \lambda
/\ell \right) ^{3}$, respectively. This is a series of powers in the
so-called Knudsen number $\mathrm{Kn}\equiv \lambda /\ell $. If $\mathrm{Kn}%
\ll 1$, the gradient expansion of $F$, Eq.\ (\ref{ExpJ}), can be truncated
at a given order and one obtains a closed macroscopic theory for the
dissipative current $J$.

\section{The role of the analytical structure of $\tilde{G}_{R}\left( 
\protect\omega \right) $}

\label{polessection}

In the previous section, it was shown how to relate the gradient expansion
of the thermodynamical force $F$ with the Taylor expansion of the Green's
function $\tilde{G}_R$. The viability of the latter required the assumption
that the singularities of $\tilde{G}_R$ are pushed to infinity by some
suitable limiting procedure. For instance, if $\tilde{F}(\omega)$ has only
support in a region of small $|\omega|$, which is well separated from the
singularities of $\tilde{G}_R$, one can devise a limiting procedure that
effectively pushes these singularities to infinity. However, a priori it is
not at all clear that $\tilde{F}(\omega)$ has vanishing support in the
region of the complex $\omega$ plane, where $\tilde{G}_R\left(\omega \right)$
has singularities.

Therefore, we have to consider the case that $\tilde{G}_R\left(\omega
\right) $ has some singularities in the complex $\omega$ plane. This fact
necessarily restricts the convergence radius of the Taylor expansion. If the
singularities are simple poles, it is better to use a Laurent expansion
around these poles.

\subsection{$\tilde{G}_{R}\left( \protect\omega \right) $ with one pole}

\begin{figure}[bht]
\hspace{-0.0cm} \includegraphics[width=6.8cm]{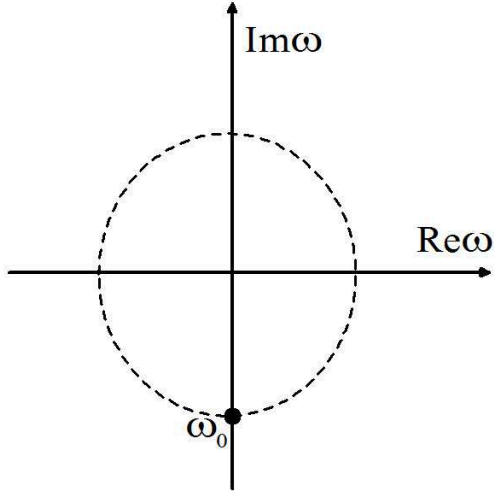} \vspace{-0.0cm} 
\vspace{-0.3cm}
\caption{{\protect\small Analytic structure of the retarded Green's function
with a singularity in $\protect\omega_0$. The dashed line illustrates the
radius of convergence of the Taylor expansion around the origin.}}
\label{OnePole}
\end{figure}

In order to illustrate this, we consider a retarded Green's function, $%
\tilde{G}_{R}\left( \omega \right) $, with a single simple pole at $\omega
_{0}$. See Fig.\ \ref{OnePole} for an illustration in the complex plane. A
function with a single pole can always be expressed in the following form, 
\begin{equation}
\tilde{G}_{R}\left( \omega \right) =\frac{f\left( \omega \right) }{\omega
-\omega _{0}}\;,  \label{Assumption1}
\end{equation}%
where $f\left( \omega \right) $ is an analytic function in the complex
plane. In order for $\mathrm{Re}\tilde{G}_{R}$ to be an even function of $%
\omega $ and $\mathrm{Im}\tilde{G}_{R}$ to be an odd function of $\omega $,
we have to require that $\omega _{0}\equiv -i\zeta $, where $\zeta $ is
positive and real, for the retarded Green's function. We also have to
require that $\mathrm{Re}\,f(\omega )$ is odd in $\omega $, while $\mathrm{Im%
}\,f(\omega )$ is even in $\omega $. Since $J$ is not a conserved quantity,
we exclude the case where the pole is at the origin, $\zeta =0$.

Since $\tilde{G}_{R}\left( \omega \right) $ has a pole at $\omega _{0}$, the
Taylor series around $\omega =0$ has a radius of convergence $\left\vert
\omega _{0}\right\vert $ and, consequently, this expansion is not able to
describe $\tilde{G}_{R}\left( \omega \right) $ beyond the pole. In such
cases, the Laurent expansion for $\tilde{G}_{R}\left( \omega \right) $around
the pole $\omega _{0}$ should be used, 
\begin{equation}
\tilde{G}_{R}\left( \omega \right) =\frac{f\left( \omega _{0}\right) }{%
\omega -\omega _{0}}+\partial _{\omega }\left. f\left( \omega \right)
\right\vert _{\omega =\omega _{0}}+\frac{1}{2}\partial _{\omega }^{2}\left.
f\left( \omega \right) \right\vert _{\omega =\omega _{0}}\left( \omega
-\omega _{0}\right) +\mathcal{O}\left[ \left( \omega -\omega _{0}\right) ^{2}%
\right] \,.  \label{Laurent}
\end{equation}

The series of positive powers in $\omega -\omega _{0}$ can be rearranged
into a series of positive powers in $\omega $. The coefficients of the
latter series can be most conveniently expressed by matching the Laurent
expansion to the Taylor expansion around $\omega =0$. To this end, we expand 
$f\left( \omega _{0}\right) \left( \omega -\omega _{0}\right) ^{-1}$ around
the origin. Then we match Eq.\ (\ref{Taylor1}) with Eq.\ (\ref{Laurent}). In
this way, we obtain e.g.\ for the coefficients of the constant and linear
terms in $\omega $ 
\begin{eqnarray}
\tilde{G}_{R}\left( 0\right) &=&-\frac{f\left( \omega _{0}\right) }{\omega
_{0}}+\partial _{\omega }\left. f\left( \omega \right) \right\vert _{\omega
=\omega _{0}}-\frac{1}{2}\partial _{\omega }^{2}\left. f\left( \omega
\right) \right\vert _{\omega =\omega _{0}}\omega _{0}+\ldots ,  \notag \\
\partial _{\omega }\left. \tilde{G}_{R}\left( \omega \right) \right\vert
_{\omega =0} &=&-\frac{f\left( \omega _{0}\right) }{\omega _{0}^{2}}+\frac{1%
}{2}\partial _{\omega }^{2}\left. f\left( \omega \right) \right\vert
_{\omega =\omega _{0}}+\ldots .  \label{trivial}
\end{eqnarray}%
We now observe that \textit{all\/} constant terms in $\omega $ in Eq.\ (\ref%
{Laurent}) can be expressed as $\tilde{G}_{R}(0)+f(\omega _{0})/\omega _{0}$%
, while \textit{all\/} linear terms in $\omega $ can be written as $\partial
_{\omega }\tilde{G}_{R}(\omega )|_{\omega =0}+f(\omega _{0})/\omega _{0}^{2}$%
. Higher-order terms in $\omega $ can be expressed in a similar fashion.
Thus, we can rewrite Eq.\ (\ref{Laurent}) in the following way%
\begin{equation}
\tilde{G}_{R}\left( \omega \right) =\frac{f\left( \omega _{0}\right) }{%
\omega -\omega _{0}}+\left[ \tilde{G}_{R}\left( 0\right) +\frac{f\left(
\omega _{0}\right) }{\omega _{0}}\right] +\left[ \partial _{\omega }\left. 
\tilde{G}_{R}\left( \omega \right) \right\vert _{\omega =0}+\frac{f\left(
\omega _{0}\right) }{\omega _{0}^{2}}\right] \omega +\mathcal{O}(\omega
^{2}).
\end{equation}%
The inverse Fourier transform of $\tilde{G}_{R}\left( \omega \right) $ is
straightforwardly obtained and has the form 
\begin{eqnarray}
G_{R}\left( t-t^{\prime }\right) &=&-if\left( \omega _{0}\right) \exp \left[
-i\omega _{0}\left( t-t^{\prime }\right) \right] \theta \left( t-t^{\prime
}\right) +\left[ \tilde{G}_{R}\left( 0\right) +\frac{f\left( \omega
_{0}\right) }{\omega _{0}}\right] \delta \left( t-t^{\prime }\right)  \notag
\\
&&+i\left[ \partial _{\omega }\left. \tilde{G}_{R}\left( \omega \right)
\right\vert _{\omega =0}+\frac{f\left( \omega _{0}\right) }{\omega _{0}^{2}}%
\right] \partial _{t}\delta \left( t-t^{\prime }\right) +\mathcal{O}%
(\partial _{t}^{2}).  \label{Triviality2}
\end{eqnarray}%
This retarded Green's function is the solution of the following differential
equation 
\begin{eqnarray}
\partial _{t}G_{R}\left( t-t^{\prime }\right) +i\omega _{0}G_{R}\left(
t-t^{\prime }\right) &=&i\omega _{0}\tilde{G}_{R}\left( 0\right) \delta
\left( t-t^{\prime }\right)  \notag \\
&-&\left[ \omega _{0}\partial _{\omega }\left. \tilde{G}_{R}\left( \omega
\right) \right\vert _{\omega =0}-\tilde{G}_{R}\left( 0\right) \right]
\partial _{t}\delta \left( t-t^{\prime }\right) +\mathcal{O}\left( \partial
_{t}^{2}\right) .  \label{Triviality3}
\end{eqnarray}%
Dividing Eq.\ (\ref{Triviality3}) by $i\omega _{0}$, multiplying by $F\left(
t^{\prime }\right) $, and integrating over $t^{\prime }$, one now obtains an 
\textit{equation of motion\/} for the current $J$ defined in Eq.\ (\ref%
{LinearRel}), instead of a simple algebraic identity as in the gradient
expansion, cf.\ Eq.\ (\ref{ExpJ}). This equation of motion reads 
\begin{equation}
\tau _{R}\partial _{t}J+J=D_{0}F+D_{1}\partial _{t}F+\mathcal{O}\left(
\partial _{t}^{2}F\right) ,  \label{RelaxJ}
\end{equation}%
where the coefficients are 
\begin{eqnarray}
\tau _{R} &=&\frac{1}{i\omega _{0}}=\frac{1}{\zeta },\text{ }  \notag \\
D_{0} &=&\tilde{G}_{R}\left( 0\right) ,\text{ }  \notag \\
D_{1} &=&i\partial _{\omega }\left. \tilde{G}_{R}\left( \omega \right)
\right\vert _{\omega =0}+D_{0}\tau _{R}=\tau _{R}\,\partial _{\omega }\left.
f\left( \omega \right) \right\vert _{\omega =0}.  \label{CoeffsA}
\end{eqnarray}%
Note that, in case $\tilde{G}_{R}(\omega )$ has a single simple pole, Eq.\ (%
\ref{RelaxJ}) is \textit{exact\/}.

It is clear that Eq.\ (\ref{RelaxJ}) is nothing but a relaxation equation
which is similar in structure to that occurring in the transient theories
for non-relativistic fluid dynamics proposed by Grad \cite{Grad} and
extended to relativistic fluids by Israel and Stewart in Ref.\ \cite{IS}.
The appearance of the time derivative of $J$ is due to the existence of the
pole in the retarded Green's function. Also, the transport coefficient $\tau
_{R}$, usually known as the relaxation time coefficient, is directly related
to the singularity, $\omega _{0}\equiv -i\zeta $, of $\tilde{G}_{R}\left(
\omega \right) $. Since $\zeta >0$, $\tau _{R}$ is real and positive, as
expected. It is interesting to note that $D_{0}\equiv \bar{D}_{0}$, with $%
\bar{D}_{0}$ from Eq.\ (\ref{Coeffs}), while $D_{1}$ is \textit{not\/}
identical to $\bar{D}_{1}$, cf.\ see Eqs.\ (\ref{Coeffs}) and (\ref{CoeffsA}%
).

It should be noted that the right-hand side of Eq.\ (\ref{RelaxJ}) can be
truncated using the same procedure employed for Eq.\ (\ref{ExpJ}). After
this truncation, the dissipative current $J$ satisfies the following
equation, 
\begin{equation}
\tau _{R}\partial _{t}J+J=D_{0}F+D_{1}\partial _{t}F\;.  \label{Foda}
\end{equation}%
As before, it was assumed that $\mathrm{Kn}\ll 1$, with $\mathrm{Kn}$ being
the appropriate Knudsen number, and terms of order $\mathcal{O}\left( 
\mathrm{Kn}^{3}\right) $ were dropped.

As was mentioned in the beginning of this section, the Taylor expansion
around $\omega =0$ is valid in a radius $\left\vert \omega _{0}\right\vert $
around the origin, see Fig.\ \ref{OnePole}. Thus, when the pole is pushed to
infinity, $\left\vert \omega _{0}\right\vert \rightarrow \infty $, the
radius of convergence of the Taylor series becomes infinite and we should
recover the gradient expansion. In fact, taking the limit $\left\vert \omega
_{0}\right\vert \rightarrow \infty $, that is, $\tau _{R}\rightarrow 0$,
cf.\ Eqs.\ (\ref{CoeffsA}), one recovers Eq.\ (\ref{ExpJ}), with \textit{%
identical\/} coefficients. The coefficient $D_{0}$ was already seen to be
identical to $\bar{D}_{0}$, while $D_{1}$ agrees with $\bar{D}_{1}$ only in
the limit of vanishing relaxation time \cite{comment1}.

In the case where the thermodynamic force varies slowly on the time scale
given by $\tau_R$, eventually, i.e., for times $t \gg \tau_R$, the
dissipative current $J$ will follow the time dependence imposed by the
right-hand side of Eq.\ (\ref{Foda}). In other words, the transient term $%
\tau_R \partial_t J$ in Eq.\ (\ref{Foda}) will become small. Then it is
permissible to replace $J$ in this term by the right-hand side of Eq.\ (\ref%
{Foda}), and we obtain up to terms of order $\mathcal{O}(\mathrm{Kn}^3)$ 
\begin{equation}
J \simeq D_0 F + D_1\partial_t F - \tau_R D_0 \partial_t F \equiv \bar{D_0}F
+ \bar{D}_1 \partial_t F\;,
\end{equation}
i.e., we recover the result (\ref{ExpJ}) given by the gradient expansion. In
this sense, the gradient expansion is the asymptotic solution of Eq.\ (\ref%
{RelaxJ}) for time $t \gg \tau_R$.

For nonrelativistic systems, when the transient dynamics can be neglected at
all times, it is known that the first-order truncation of the gradient
expansion can actually serve not only as an asymptotic solution, but as an
effective theory to describe the system, e.g., substituting the first-order
result into the conservation equations one obtains the nonrelativistic
diffusion equation and the nonrelativistic Navier-Stokes equation. However,
for relativistic theories this is not possible because of the violation of
causality. As was mentioned in the Introduction, because of Eq.\ (\ref{asc})
for any non-zero value of the transport coefficients (shear viscosity, bulk
viscosity, etc.) it is not possible to take the (acausal) limit of vanishing
relaxation time, if one wants to obtain causal and stable relativistic
fluid-dynamical equations of motion.

\subsection{$\tilde{G}_{R}\left( \protect\omega \right) $ with two poles}

In order to better understand the consequences induced by the retarded
Green's function's nontrivial analytic structure, it is useful to analyze in
detail the case where $\tilde{G}_{R}\left( \omega \right) $ has two poles, $%
\omega _{1}$ and $\omega _{2}$,%
\begin{equation}
\tilde{G}_{R}\left( \omega \right) =\frac{f_{1}\left( \omega \right) }{%
\omega -\omega _{1}}+\frac{f_{2}\left( \omega \right) }{\omega -\omega _{2}}.
\end{equation}%
We employ exactly the same steps as before and expand \textit{each} term of $%
\tilde{G}_{R}\left( \omega \right) $ in a Laurent series around its
respective pole. The result is%
\begin{eqnarray}
\tilde{G}_{R}\left( \omega \right) &=&\frac{f_{1}\left( \omega _{1}\right) }{%
\omega -\omega _{1}}+\frac{f_{2}\left( \omega _{2}\right) }{\omega -\omega
_{2}}+\partial _{\omega }\left. f_{1}\left( \omega \right) \right\vert
_{\omega =\omega _{1}}+\partial _{\omega }\left. f_{2}\left( \omega \right)
\right\vert _{\omega =\omega _{2}}  \notag \\
&&+\frac{1}{2}\partial _{\omega }^{2}\left. f_{1}\left( \omega \right)
\right\vert _{\omega =\omega _{1}}\left( \omega -\omega _{1}\right) +\frac{1%
}{2}\partial _{\omega }^{2}\left. f_{2}\left( \omega \right) \right\vert
_{\omega =\omega _{2}}\left( \omega -\omega _{2}\right) 
 +\mathcal{O}\left[(\omega -\omega _{i})^{2}\right] \,.
\end{eqnarray}%
As before, we can match this expansion to the Taylor expansion near the
origin. This enables us to rewrite $\tilde{G}_{R}\left( \omega \right) $ as 
\begin{eqnarray}
\tilde{G}_{R}\left( \omega \right) = \frac{f_{1}\left( \omega _{1}\right) }{%
\omega -\omega _{1}}+\frac{f_{2}\left( \omega _{2}\right) }{\omega -\omega
_{2}}& +& \left[ \tilde{G}_{R}\left( 0\right) +\frac{f_{1}\left( \omega
_{1}\right) }{\omega _{1}}+\frac{f_{2}\left( \omega _{2}\right) }{\omega _{2}%
}\right]  \notag \\
&+ &\left[ \partial _{\omega }\left. \tilde{G}_{R}\left( \omega \right)
\right\vert _{\omega =0}+\frac{f_{1}\left( \omega _{1}\right) }{\omega
_{1}^{2}}+\frac{f_{2}\left( \omega _{2}\right) }{\omega _{2}^{2}}\right]
\omega \notag \\
& + & \left[ \frac{1}{2}\left.\partial_\omega^2\, \tilde{G}_R(\omega)
\right|_{\omega=0} + \frac{f_1(\omega_1)}{\omega_1^3}
+ \frac{f_2(\omega_2)}{\omega_2^3} \right]\, \omega^2 
+\mathcal{O}(\omega ^{3}).
\end{eqnarray}%
With the last expression, we can determine the Green's function $%
G_{R}(t-t^{\prime })$ and also its equation of motion. Similarly to the
previous section, it is straightforward to show that%
\begin{eqnarray}
G_{R}\left( t-t^{\prime }\right) & =& -i\left\{ \frac{{}}{{}}f_{1}\left(
\omega _{1}\right) \exp \left[ -i\omega _{1}\left( t-t^{\prime }\right) %
\right] +f_{2}\left( \omega _{2}\right) \exp \left[ -i\omega _{2}\left(
t-t^{\prime }\right) \right] \right\} \theta \left( t-t^{\prime }\right) 
\notag \\
& &+\left[ \tilde{G}_{R}\left( 0\right) +\frac{f_{1}\left( \omega _{1}\right) 
}{\omega _{1}}+\frac{f_{2}\left( \omega _{2}\right) }{\omega _{2}}\right]
\delta \left( t-t^{\prime }\right)  \notag \\
& & + i\left[ \partial _{\omega }\left. \tilde{G}_{R}\left( \omega \right)
\right\vert _{\omega =0}+\frac{f_{1}\left( \omega _{1}\right) }{\omega
_{1}^{2}}+\frac{f_{2}\left( \omega _{2}\right) }{\omega _{2}^{2}}\right]
\partial _{t}\delta \left( t-t^{\prime }\right) \notag \\
& & - \left[ \frac{1}{2}\,\partial_\omega^2\left. \tilde{G}_R(\omega)
\right|_{\omega =0}+\frac{f_1(\omega_1)}{\omega_1^3}
+\frac{f_2(\omega_2)}{\omega_2^3}\right]\,
\partial_t^2\,\delta ( t-t') +\mathcal{O}(\partial_{t}^{3})\,. 
\end{eqnarray}%
The equation satisfied by $G_{R}$ is%
\begin{eqnarray}
\lefteqn{-\partial _{t}^{2}G_{R}\left( t-t^{\prime }\right) =i\left( \omega
_{1}+\omega _{2}\right) \partial _{t}G_{R}\left( t-t^{\prime }\right)
-\omega _{1}\omega _{2}G_{R}\left( t-t^{\prime }\right) +\omega _{1}\omega
_{2}\tilde{G}_{R}\left( 0\right) \delta \left( t-t^{\prime }\right)}  \notag
\\
&+&i\left[ \omega _{1}\omega _{2}\partial _{\omega }\left. \tilde{G}%
_{R}\left( \omega \right) \right\vert _{\omega =0}-\left( \omega _{1}+\omega
_{2}\right) \tilde{G}_{R}\left( 0\right) \right] \partial _{t}\delta \left(
t-t^{\prime }\right) \notag \\
&-&\left[ \frac{1}{2}\, \omega_1\,\omega_2\, 
\partial_\omega^2\left. \tilde{G}_R(\omega)
\right|_{\omega =0} -
\left( \omega_1+\omega_2 \right)\,
\partial_\omega \left. \tilde{G}_R(\omega) \right|_{\omega=0}
+ \tilde{G}_R(0) \right]
\, \partial_t^2\,\delta(t-t') +\mathcal{O}(\partial _{t}^{3})\;,\;\;\;\;
\end{eqnarray}%
The main difference to the previous case is that, due to the existence of a
second pole, the Green's function now satisfies a \textit{second}-order
differential equation, instead of a first-order one. As will be shown later,
the order of the differential equation satisfied by $\tilde{G}_{R}\left(
\omega \right) $ is equal to the number of its singularities. Dividing by $%
-\omega _{1}\omega _{2}$, multiplying the equation by $F\left( t^{\prime
}\right) $, and integrating over $t^{\prime }$ we can determine the equation
of motion for $J$, 
\begin{equation}
\chi _{2}\partial _{t}^{2}J+\chi _{1}\partial _{t}J+J=D_{0}F\left( t\right)
+D_{1}\partial _{t}F\left( t\right) +D_{2}\partial _{t}^{2}F\left( t\right) +%
\mathcal{O}\left[ \left( \lambda /\ell \right) ^{4}\right] .  \label{GreatEq}
\end{equation}%
We introduced the following transport coefficients,%
\begin{eqnarray}
\chi _{2} &=&-\frac{1}{\omega _{1}\omega _{2}},  \notag \\
\chi _{1} &=&\frac{1}{i\omega _{1}}+\frac{1}{i\omega _{2}},  \notag \\
D_{0} &=&\tilde{G}_{R}\left( 0\right) ,\text{ }  \notag \\
D_{1} &=&i\partial _{\omega }\left. \tilde{G}_{R}\left( \omega \right)
\right\vert _{\omega =0}+D_{0}\chi _{1}\,,  \notag \\
D_{2} &=&-\frac{1}{2}\partial _{\omega }^{2}\left. \tilde{G}_{R}\left(
\omega \right) \right\vert _{\omega =0}+D_{1}\chi _{1}+D_{0}\left( \chi
_{2}- \chi _{1}^{2}\right) ,  \label{Coeffs2}
\end{eqnarray}%
Note that $\chi _{2}$ and $\chi _{1}$ have contributions from \textit{both}
poles.

\begin{figure}[tbp]
\begin{minipage}{.45\linewidth}
\includegraphics[width=6.0cm]{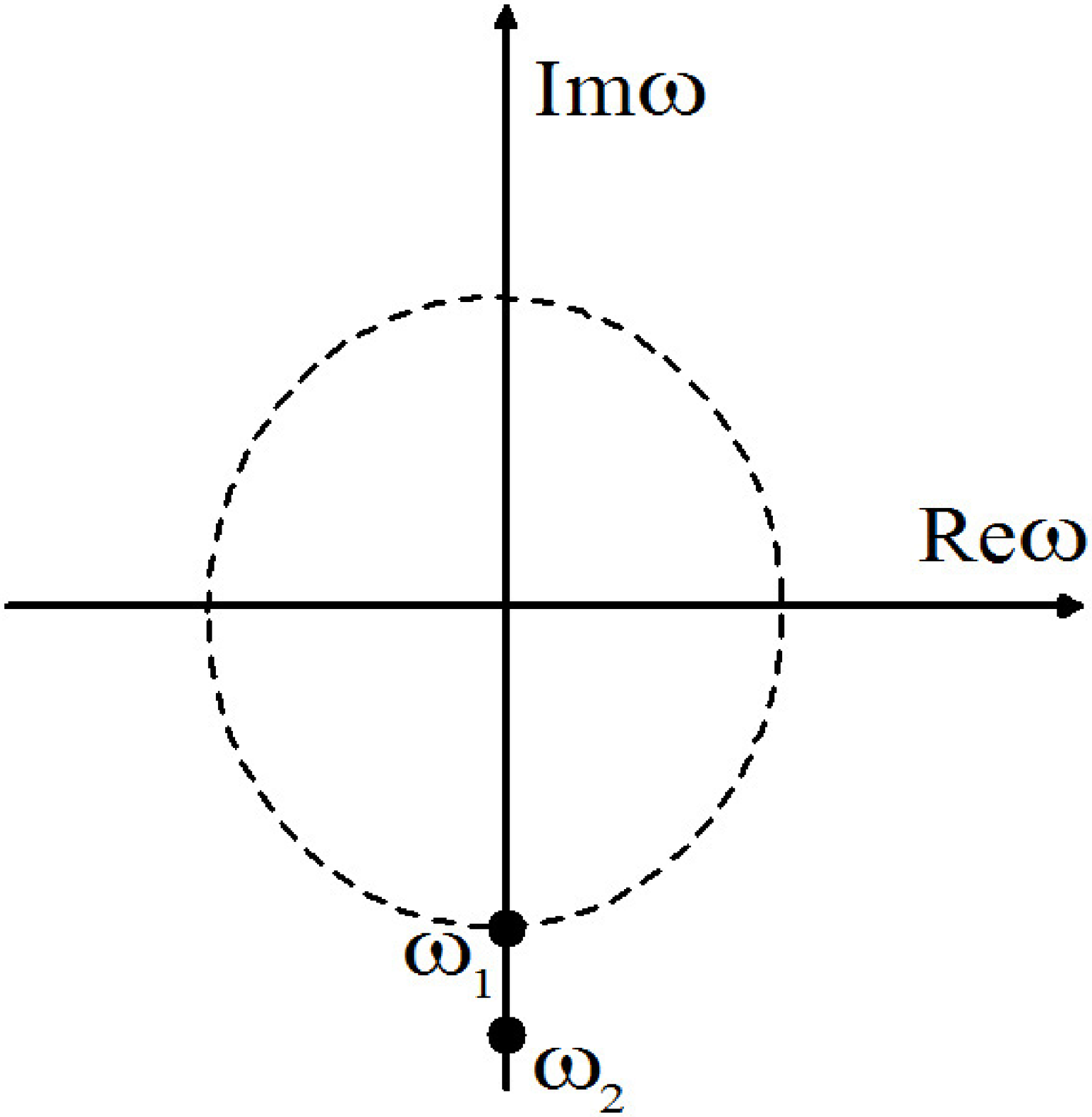} 
\end{minipage}  
\begin{minipage}{.45\linewidth}
\includegraphics[width=6.2cm]{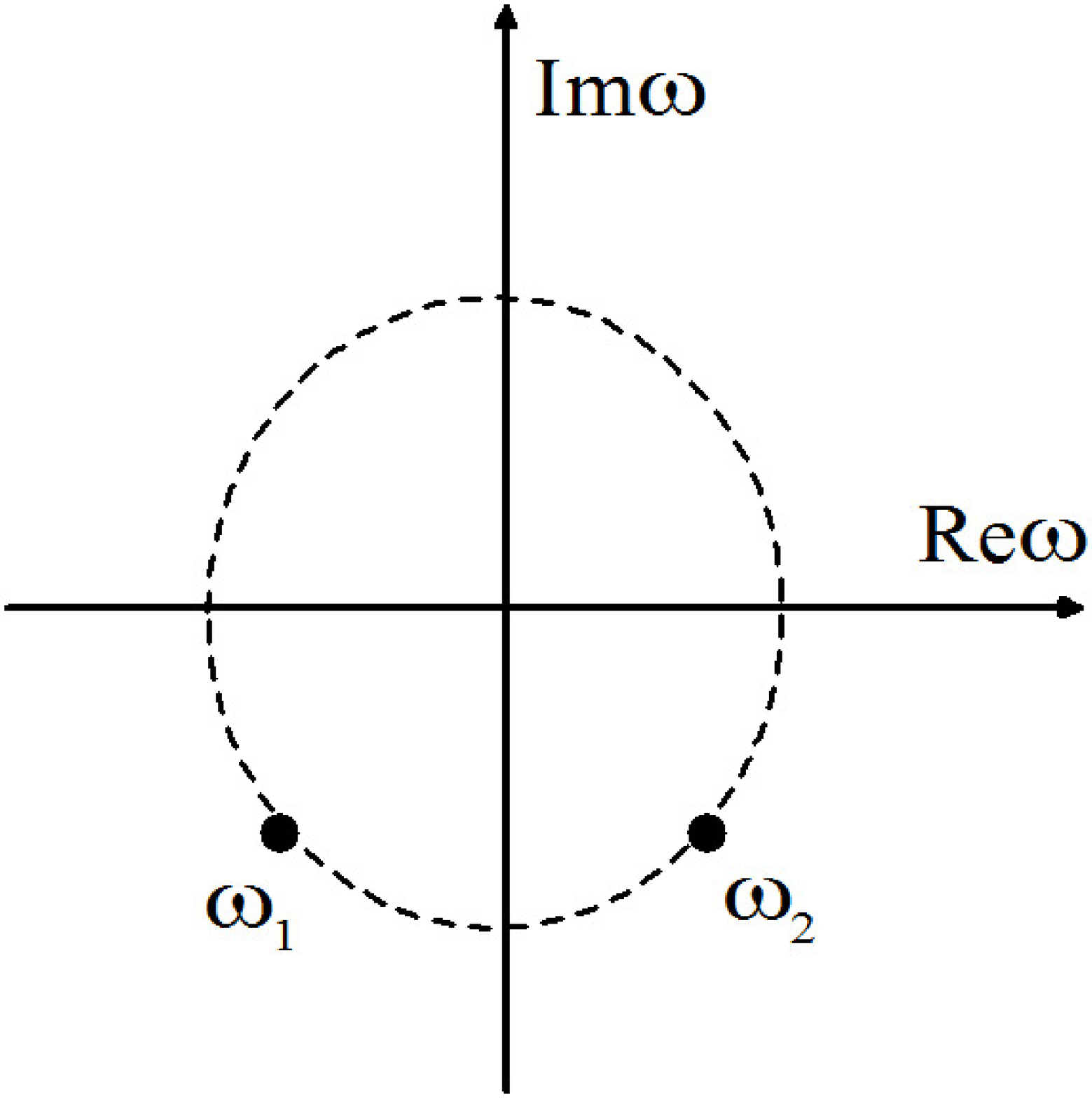} 
\end{minipage} 
\caption{Analytic structure of the retarded Green's function with two
singularities, $\protect\omega_1$ and $\protect\omega_2$. Case (i), in which
both poles are on the imaginary axis, is illustrated on the left. Case (ii),
in which both poles are symmetric around the imaginary axis, is illustrated
on the right. The dashed line illustrates the radius of convergence of the
Taylor expansion around the origin.}
\label{TwoPoles}
\end{figure}

Next, we shall investigate under which circumstances a relaxation equation
for $J$ can be obtained. Due to time reversal invariance, the two poles of $%
\tilde{G}_{R}\left( \omega \right) $ can appear in two ways: (i) both poles
are on the imaginary axis, in which case we assume, without any loss of
generality, that $\left\vert \omega _{2}\right\vert >\left\vert \omega
_{1}\right\vert $; (ii) both poles have the same imaginary part, but
opposite real parts, being symmetric with respect to the imaginary axis. In
this case, $\left\vert \omega _{1}\right\vert =\left\vert \omega
_{2}\right\vert $. See Fig.\ \ref{TwoPoles} for details. Both cases reflect
distinct physical scenarios.

Let us consider the case in which the thermodynamic force is turned off, and
the current $J$ is left to relax to equilibrium. This process is governed by
the equation%
\begin{equation}
\chi _{2}\partial _{t}^{2}J+\chi _{1}\partial _{t}J+J=0.  \label{2ndorderJ}
\end{equation}%
This is the equation of motion of a damped harmonic oscillator, 
\begin{equation}
\ddot{x}+2\,\gamma \,\dot{x}+\omega _{0}^{2}\,x=0\;.
\end{equation}%
The harmonic oscillator is overdamped, if $\gamma >\omega _{0}$, and
underdamped, if $\gamma <\omega _{0}$. Identifying the coefficients, we find 
\begin{equation}
\gamma =\frac{\chi _{1}}{2\,\chi _{2}}\;,\;\;\;\;\omega _{0}^{2}=\frac{1}{%
\chi _{2}}\;.
\end{equation}%
Therefore, if $\chi _{1}^{2}>4\,\chi _{2}$, the dissipative
current $J$ relaxes to equilibrium without oscillating, while for $
\chi _{1}^{2}<4\,\chi _{2}$, it relaxes in an oscillatory fashion.
Using the definitions (\ref{Coeffs2}), we see that in the overdamped case, $%
4\omega _{1}\omega _{2}>(\omega _{1}+\omega _{2})^{2}$, while in the
underdamped case, $4\omega _{1}\omega _{2}<(\omega _{1}+\omega _{2})^{2}$.

In case (i), both poles are purely imaginary, $\omega _{i}\equiv -i\zeta
_{i} $, with $\zeta _{i}>0$, $i=1,2$. Since always $4\zeta _{1}\zeta
_{2}<(\zeta _{1}+\zeta _{2})^{2}$, we are in the overdamped limit. In this
case, a relaxation equation is obtained by applying the limiting procedure $%
\,\chi _{2}\rightarrow 0$. That is, the second pole is pushed to infinity,
in which case the relaxation time is given by the inverse of the first pole, 
\begin{equation*}
\tau _{R}\equiv \chi _{1}=\frac{1}{i\omega _{1}}.
\end{equation*}%
Naturally, this theory is only valid for frequencies that are small compared
to the second pole.

In case (ii), $\left\vert \omega _{1}\right\vert =\left\vert \omega
_{2}\right\vert $ and $\omega _{1}=-\omega _{2}^{\ast }$. We always have $%
|\omega _{1}|^{2}>(\mathrm{Im}\omega _{1})^{2}$, which can be rewritten in
the form $4\omega _{1}\omega _{2}=-4|\omega _{1}|^{2}<(\omega _{1}-\omega
_{1}^{\ast })^{2}=(\omega _{1}+\omega _{2})^{2}$. Thus, we are in the
underdamped limit. In this case, due to the symmetries of the retarded
Green's function, one \textit{cannot\/} disregard one of the poles while
keeping the other. Thus, the term including the second time derivative of $J$
must be included (otherwise there would be no oscillation) and the full
equation of motion must be solved. It is important to remark that, in this
case, the coefficient $\chi _{1}$ cannot be interpreted as a relaxation
time. 

Note that the exponential decay of $J(t)$ in case (i) 
can also be seen when inserting the Green's function
$\tilde{G}_R(\omega)$ into Eq.\ (\ref{6}) and performing the
$\omega$ integral via contour integration, picking up the poles via
the residue theorem. In case (ii), one gets an exponentially
damped factor from the imaginary part of the pole, 
$\sim \exp(-|{\rm Im} \omega_i|\,t)$, and an oscillatory factor from
the real part, $\sim \exp(i\,{\rm Re}\omega_i\,t)$. If $|{\rm Im}\omega_i|
\gg |{\rm Re}\omega_i|$, the damping is much stronger than the
oscillation, already within a single oscillation, the dissipative
current has decayed a couple of $e-$folds towards its stationary
solution. On the other hand, if  $|{\rm Im}\omega_i|
\ll |{\rm Re}\omega_i|$, the current will oscillate many times before
a substantial decay occurs.
The oscillatory behavior of the dissipative current in case (ii)
was already noticed in Ref.\ \cite{Kovtun:2005ev}.

It is clear from this analysis that the derivation of relaxation equations
for systems with more than one pole is not possible if the first pole does
not lie on the imaginary axis. With this in mind, we can finally consider
the general case of an arbitrary number of poles. The result will be
qualitatively similar to what was found above.

\subsection{ $\tilde{G}_{R}\left( Q\right) $ with $N$ poles}

Now we assume that $\tilde{G}_{R}\left( Q\right) $ has $N$ poles in the
complex $\omega $-plane, $\omega _{1}\left( \mathbf{q}\right) $, \ldots\ , $%
\omega _{N}\left( \mathbf{q}\right) $. We furthermore restore the $\mathbf{q}
$ dependence which was neglected in the previous sections, and assume that
the Green's function is analytic in $\mathbf{q}$. Since $J$ is not a
conserved quantity, we can safely assume that all poles remain within a
finite distance from the origin even when $\mathbf{q}\rightarrow 0$. Here we
consider the case of a finite (but arbitrarily large) number of poles. Also,
in many cases $\tilde{G}_{R}\left( Q\right) $ contains branch cuts. We
assume that these branch cuts are located further from the origin of the
complex frequency plane than these $N$ poles and thus will not be considered
any further in this work. The retarded Green's function can be written in
the general form%
\begin{equation}
\tilde{G}_{R}\left( Q\right) =\sum_{i=1}^{N}\frac{f_{i}\left( Q\right) }{%
\omega -\omega _{i}\left( \mathbf{q}\right) }=\frac{\Xi \left( Q\right) }{%
\left[ \omega -\omega _{1}\left( \mathbf{q}\right) \right] \cdots \left[
\omega -\omega _{N}\left( \mathbf{q}\right) \right] },
\end{equation}%
where the $f_{i}\left( Q\right) $ and $\Xi \left( Q\right) $ are analytic
functions in the complex $\omega $ plane. We now use the identity%
\begin{equation}
1+\Phi _{1}\left( \mathbf{q}\right) \left( -i\omega \right) +\ldots +\Phi
_{N}\left( \mathbf{q}\right) \left( -i\omega \right) ^{N}=\left( -1\right)
^{N}\frac{\left[ \omega -\omega _{1}\left( \mathbf{q}\right) \right] \cdots %
\left[ \omega -\omega _{N}\left( \mathbf{q}\right) \right] }{\omega
_{1}\left( \mathbf{q}\right) \cdots \omega _{N}\left( \mathbf{q}\right) }\;,
\end{equation}%
where 
\begin{equation}
\Phi _{m}\left( \mathbf{q}\right) =\left( -i\right) ^{m}\sum_{1\leq
i_{1}...<i_{m}\leq N}\frac{1}{\omega _{i_{1}}\left( \mathbf{q}\right) \cdots
\omega _{i_{m}}\left( \mathbf{q}\right) }\;.
\end{equation}%
From this expression, we can find the following equation for $\tilde{G}%
_{R}\left( Q\right) $, 
\begin{equation}
\left[ 1+\Phi _{1}\left( \mathbf{q}\right) \left( -i\omega \right) +\ldots
+\Phi _{N}\left( \mathbf{q}\right) \left( -i\omega \right) ^{N}\right] 
\tilde{G}_{R}\left( Q\right) =\frac{\left( -1\right) ^{N}\Xi \left( Q\right) 
}{\omega _{1}\left( \mathbf{q}\right) \cdots \omega _{N}\left( \mathbf{q}%
\right) }\;.
\end{equation}%
After taking the Fourier transform and expanding the functions $\Phi
_{m}\left( \mathbf{q}\right) $ in a Taylor series around $\mathbf{q}=0$, we
obtain a differential equation satisfied by $G_{R}$. It is clear that this
will be a linear differential equation of order $N$ in time. The equation of
motion for $J$ can be obtained in the same way as before. The result is%
\begin{equation}
\chi _{N}\partial _{t}^{N}J+\ldots +\chi _{1}\partial _{t}J+J=D_{0}F+\ldots
+D_{N}\partial _{t}^{N}F+\mathcal{O}\left( \partial _{t}^{N+1}F,\partial _{%
\mathbf{x}}\right) .  \label{Sugoy}
\end{equation}%
Here, we omitted all the terms involving spatial derivatives. The
coefficients $\chi _{m}$ are 
\begin{equation}
\chi _{m}=\Phi _{m}\left( \mathbf{0}\right) =\left( -i\right)
^{m}\sum_{1\leq i_{1}...<i_{m}\leq N}\frac{1}{\omega _{i_{1}}\left( \mathbf{0%
}\right) \cdots \omega _{i_{m}}\left( \mathbf{0}\right) }\;.
\label{very nice}
\end{equation}%
As before, the coefficients $D_{0}$, $D_{1}$ and $D_{2}$ can be expressed as 
\begin{eqnarray}
D_{0} &=&\left. \tilde{G}_{R}\left( \omega ,\mathbf{0}\right) \right\vert
_{\omega =0}\;,  \notag \\
D_{1} &=&i\partial _{\omega }\left. \tilde{G}_{R}\left( \omega ,\mathbf{0}%
\right) \right\vert _{\omega =0}+D_{0}\chi _{1}\;,  \notag \\
D_{2} &=&-\frac{1}{2}\partial _{\omega }^{2}\left. \tilde{G}_{R}\left(
\omega ,\mathbf{0}\right) \right\vert _{\omega =0}+D_{1}\chi _{1}+D_0\left(
\chi _{2}-\chi_{1}^{2}\right)\;,  \label{very nice3}
\end{eqnarray}%
cf.\ Eq.\ (\ref{Coeffs2}). In general, the coefficients $D_{k}$ have the
following form%
\begin{equation}
D_{k}=i^{k}\frac{\left( -1\right) ^{N}}{k!}\frac{\partial _{\omega
}^{k}\left. \Xi \left( \omega ,\mathbf{0}\right) \right\vert _{\omega =0}}{%
\omega _{1}\left( \mathbf{0}\right) \cdots \omega _{N}\left( \mathbf{0}%
\right) }.
\end{equation}

We remark that up to now we have not employed any approximation besides the
initial assumptions regarding the singularities of $\tilde{G}_{R}$. As
mentioned before, if there is a clear separation of scales, it is possible
to simplify the equation of motion. Since $D_{n}\sim \lambda ^{n+1}$,
assuming that $\lambda /\ell \ll 1$, we can safely assume $D_{0}F\gg
D_{1}\partial _{t}F\gg D_{2}\partial _{t}^{2}F$ and truncate the right-hand
side of Eq.\ (\ref{Sugoy}).

As discussed in the previous section, a relaxation equation can be obtained
only for the cases in which the pole nearest to the origin (in the following
referred to as the ``first pole'') lies on the imaginary axis. In this case,
the limiting procedure which pushes all the other poles to infinity, $\chi
_{i}\rightarrow 0$, $i\geq 2$, can be applied without breaking any
symmetries of the retarded Green's function. Assuming that this can be done,
we obtain the following equation of motion for $J$%
\begin{equation}
\tau _{R}\partial _{t}J+J=D_{0}F+D_{1}\partial _{t}F+D_{2}\partial _{t}^{2}F+%
\mathcal{O}\left( D_{3}\partial _{t}^{3}F,\partial _{\mathbf{x}}\right) \;.
\label{38}
\end{equation}%
where 
\begin{eqnarray}
\tau _{R} &=&\frac{1}{i\omega _{1}\left( \mathbf{0}\right) },  \notag \\
D_{0} &=&\left. \tilde{G}_{R}\left( \omega ,\mathbf{0}\right) \right\vert
_{\omega =0}\;,  \notag \\
D_{1} &=&i\partial _{\omega }\left. \tilde{G}_{R}\left( \omega ,\mathbf{0}%
\right) \right\vert _{\omega =0}+D_{0}\,\tau _{R}\;,  \notag \\
D_{2} &=&-\frac{1}{2}\partial _{\omega }^{2}\left. \tilde{G}_{R}\left(
\omega ,\mathbf{0}\right) \right\vert _{\omega =0}+D_{1}\,\tau _{R}-D_0\,
\tau _{R}^{2}\;,  \label{very nice4}
\end{eqnarray}
On the other hand, if the first pole and, for reasons of symmetry, its
counterpart on the other side of the imaginary axis, have nonzero real
parts, the dissipative current will oscillate and the equation of motion
cannot be reduced to a simple relaxation equation, even in the
small-frequency domain.

\section{Applications}

\subsection{The Linearized Boltzmann Equation}

\label{Boltzmannsection}

The discussion presented above is valid for both weakly and strongly coupled
theories. The prime example of a weakly coupled theory is given by the
Boltzmann equation. In this section we calculate the shear viscosity and
relaxation time coefficients for a weakly coupled gas via the Boltzmann
equation following the general method presented above.

We start from the relativistic Boltzmann equation%
\begin{equation}
K\cdot \partial f_{\mathbf{k}}=C\left[ f\right] ,  \label{BE}
\end{equation}%
where $K=(k_0,\mathbf{k})$, $k_0=\sqrt{\mathbf{k}^{2}+m^{2}}$, and $m$ is
the particle mass. We use the notation $f_{\mathbf{k}}\left( X\right)
=f\left( X,K\right) $. We consider the linearized Boltzmann equation
around the classical equilibrium state, $f_{0\mathbf{k}}\equiv $ $\exp
\left( y_{0\mathbf{k}}\right) $, where $y_{0\mathbf{k}}=\alpha _{0}-\beta
_{0}E_{\mathbf{k}}$, with the inverse temperature $\beta_{0}$, the ratio of
the chemical potential to temperature $\alpha_0$, and the energy in the
local rest frame $E_{\mathbf{k}} \equiv U \cdot K$, respectively. Here, $%
u^\mu$ is the fluid four-velocity.

The Boltzmann equation can be written as%
\begin{equation}
\delta \dot{f}_{\mathbf{k}}+E_{\mathbf{k}}^{-1}K\cdot \nabla \delta f_{%
\mathbf{k}}-\hat{C}\left( X,K\right) \delta f_{\mathbf{k}}=\mathcal{S}\left(
X,K\right) ,  \label{LBE1}
\end{equation}%
where we defined $\delta f_{\mathbf{k}}\equiv f_{\mathbf{k}}-f_{0\mathbf{k}}$%
, $\dot{A}\equiv U\cdot \partial A$, $\nabla _{\mu }\equiv \Delta _{\mu \nu
}\partial ^{\nu }$, with $\Delta ^{\mu \nu }=\eta^{\mu \nu }-u^{\mu }u^{\nu }$
being the 3-space projector orthogonal to $u^{\mu }$, and%
\begin{eqnarray}
\mathcal{S}\left( X,K\right) &\equiv &f_{0\mathbf{k}}\left[ -\dot{\alpha}%
_{0}+E_{\mathbf{k}}\dot{\beta}_{0}+K\cdot \left( \beta _{0}\dot{U}+\nabla
\beta _{0}-E_{\mathbf{k}}^{-1}\nabla \alpha _{0}\right) \right.  \notag \\
&&\left. +\frac{\beta _{0}}{3}E_{\mathbf{k}}^{-1}\left( m^{2}-E_{\mathbf{k}%
}^{2}\right) \theta +\beta _{0}E_{\mathbf{k}}^{-1}k^{\left\langle \mu
\right. }k^{\left. \nu \right\rangle }\sigma _{\mu \nu }\right] .
\end{eqnarray}%
Here, we introduced the expansion scalar $\theta \equiv \partial \cdot U$
and the shear tensor $\sigma ^{\mu \nu }\equiv \nabla ^{\left\langle \mu
\right. }u^{\left. \nu \right\rangle }$, where $A^{\left\langle \mu \nu
\right\rangle }=\Delta ^{\mu \nu \alpha \beta }A_{\alpha \beta }$ with $%
\Delta ^{\mu \nu \alpha \beta }=\left( \Delta ^{\mu \alpha }\Delta ^{\nu
\beta }+\Delta ^{\mu \beta }\Delta ^{\nu \alpha }\right) /2-\Delta ^{\mu \nu
}\Delta ^{\alpha \beta }/3$ being the double symmetric traceless projection
operator. In Eq.\ (\ref{LBE1}), we also introduced the collision operator 
\begin{equation}
\hat{C}\left( X,K\right) \delta f_{\mathbf{k}}=\frac{1}{\nu E_{\mathbf{k}}}%
\int dK^{\prime }dPdP^{\prime }W_{\mathbf{kk}\prime \rightarrow \mathbf{pp}%
\prime }f_{0\mathbf{k}}f_{0\mathbf{k}^{\prime }}\left( \frac{\delta f_{%
\mathbf{p}^{\prime }}}{f_{0\mathbf{p}^{\prime }}}+\frac{\delta f_{\mathbf{p}}%
}{f_{0\mathbf{p}}}-\frac{\delta f_{\mathbf{k}^{\prime }}}{f_{0\mathbf{k}%
^{\prime }}}-\frac{\delta f_{\mathbf{k}}}{f_{0\mathbf{k}}}\right) ,
\end{equation}%
where $W_{\mathbf{kk}\prime \rightarrow \mathbf{pp}\prime }$ is the
transition rate, $\nu $ is the symmetry factor ($\nu =2$ for identical
particles) and 
\begin{equation}
dK=\frac{d^{3}\mathbf{k}}{\left( 2\pi \right) ^{3}E_{\mathbf{k}}}.
\end{equation}

We now consider Eq.\ (\ref{LBE1}) in the local rest frame, where $%
U=(1,0,0,0) $. We assume a situation where the fluid does not accelerate, $%
\dot{U} \equiv 0$, and does not expand, $\theta \equiv 0$, and where
temperature and chemical potential are constant. With these assumptions the
source term reduces to%
\begin{equation}  \label{44}
\mathcal{S}\left( X,K\right) =\beta _{0}f_{0\mathbf{k}}E_{\mathbf{k}%
}^{-1}k^{\left\langle \mu \right. }k^{\left. \nu \right\rangle }\sigma _{\mu
\nu }\left( X\right)\; ,
\end{equation}
and the collision operator no longer depends on $X$, $\hat{C}(X,K) \equiv 
\hat{C}(K)$. The Boltzmann equation (\ref{LBE1}) takes the form 
\begin{equation}  \label{LBE}
\partial_t \delta f_{\mathbf{k}}+ \mathbf{v} \cdot \nabla \delta f_{\mathbf{k%
}}-\hat{C}\left( K\right) \delta f_{\mathbf{k}}=\mathcal{S}\left(
X,K\right)\;,
\end{equation}
where $\mathbf{v} \equiv \mathbf{k}/E_{\mathbf{k}}$.

We solve the inhomogeneous, linear, partial integro-differential equation (%
\ref{LBE}) for $\delta f_{\mathbf{k}}$ in four-momentum space. Taking the
Fourier transform we obtain 
\begin{equation}
-i\omega \delta \tilde{f}_{\mathbf{k}}\left( Q\right) +i\mathbf{v}\cdot 
\mathbf{q}\,\delta \tilde{f}_{\mathbf{k}}\left( Q\right) -\hat{C}\left(
K\right) \delta \tilde{f}_{\mathbf{k}}\left( Q\right) =\mathcal{\tilde{S}}%
\left( Q,K\right) \,.  \label{LBEF}
\end{equation}%
From now on, it is important not to confuse the momenta of the particles, $%
K $, $K^{\prime }$, $P$, and $P^{\prime }$, with the variable $Q$ from the
Fourier transformation.

The formal solution of Eq.\ (\ref{LBEF}) can be expressed in the following
form%
\begin{equation}
\delta \tilde{f}_{\mathbf{k}}\left( Q\right) =\frac{1}{-i\omega +i\mathbf{v}%
\cdot \mathbf{q}-\hat{C}\left( K\right) }\mathcal{\tilde{S}}\left(
Q,K\right) .  \label{Sol}
\end{equation}%
Note that the shear stress tensor, $\pi ^{\mu \nu }$, can be expressed in
terms of $\delta f_{\mathbf{k}}$ as \cite{dkr}%
\begin{equation}
\pi ^{\mu \nu }=\int dK\,k^{\left\langle \mu \right. }k^{\left. \nu
\right\rangle }\delta f_{\mathbf{k}}.  \label{shear}
\end{equation}%
Taking the Fourier transform of Eq.\ (\ref{shear}) and using Eqs.\ (\ref{44}%
) and (\ref{Sol}), we obtain 
\begin{equation}
\tilde{\pi}^{\mu \nu }\left( Q\right) =\int dK\, k^{\left\langle \mu \right.
}k^{\left. \nu \right\rangle }\frac{1}{-i\omega +i\mathbf{v}\cdot \mathbf{q}-%
\hat{C}\left( K\right) }\beta _{0}f_{0\mathbf{k}}E_{\mathbf{k}%
}^{-1}k^{\left\langle \alpha \right. }k^{\left. \beta \right\rangle }\tilde{%
\sigma}_{\alpha \beta }\left( Q\right) .  \label{EqWow}
\end{equation}

Note that Eq.\ (\ref{EqWow}) has the following form, 
\begin{equation}
\tilde{\pi}^{\mu \nu }\left( Q\right) =\tilde{G}_{R}^{\mu \nu \alpha \beta
}\left( Q\right) \tilde{\sigma}_{\alpha \beta }\left( Q\right) ,
\label{LinBoltz}
\end{equation}%
where we introduced 
\begin{equation}
\tilde{G}_{R}^{\mu \nu \alpha \beta }\left( Q\right) =\int
dK\,k^{\left\langle \mu \right. }k^{\left. \nu \right\rangle }\frac{1}{%
-i\omega +i\mathbf{v}\cdot \mathbf{q}-\hat{C}\left( K\right) }\beta _{0}f_{0%
\mathbf{k}}E_{\mathbf{k}}^{-1}k^{\left\langle \alpha \right. }k^{\left.
\beta \right\rangle }.  \label{oh_man}
\end{equation}

As already mentioned, we want to calculate the two main transport
coefficients that describe the linearized dynamics of the shear stress
tensor: the shear relaxation time, $\tau _{\pi }$, and the shear viscosity
coefficient, $\eta $. Let us start by defining the function 
\begin{equation}
B^{\alpha \beta }=\frac{1}{-i\omega +i\mathbf{v}\cdot \mathbf{q}-\hat{C}%
\left( K\right) }\beta _{0}f_{0\mathbf{k}}E_{\mathbf{k}}^{-1}k^{\left\langle
\alpha \right. }k^{\left. \beta \right\rangle },
\end{equation}%
where, by definition, $B^{\alpha \beta }$ satisfies 
\begin{equation}
\left[ -i\omega +i\mathbf{v}\cdot \mathbf{q}-\hat{C}\left( K\right) \right]
B^{\alpha \beta }\left( Q,K\right) =\beta _{0}E_{\mathbf{k}%
}^{-1}k^{\left\langle \alpha \right. }k^{\left. \beta \right\rangle }f_{0%
\mathbf{k}}.  \label{equation}
\end{equation}%
In general, $B^{\alpha \beta }$ is a function of $Q$ and $K$. However, from
Eqs.\ (\ref{very nice}) and (\ref{very nice3}) we already know that, in
order to calculate the relaxation time and the viscosity coefficient, it is
sufficient to consider the case $\mathbf{q}=0$. Then $B^{\alpha \beta
}=B^{\alpha \beta }\left( \omega ,K\right) $. The dependence of $B^{\alpha
\beta }$ on $K$ can be expressed via the following expansion, 
\begin{equation}
B^{\alpha \beta }\left( \omega ,K\right) =f_{0\mathbf{k}}k^{\left\langle
\alpha \right. }k^{\left. \beta \right\rangle }\sum_{n=0}^{\infty
}a_{n}\left( \omega \right) E_{\mathbf{k}}^{n}.  \label{expansion}
\end{equation}%
Substituting Eq.\ (\ref{expansion}) into Eq.\ (\ref{oh_man}), it follows
that 
\begin{eqnarray}
\tilde{G}_{R}^{\mu \nu \alpha \beta }\left( \omega ,\mathbf{0}\right)
&=&\sum_{n=0}^{\infty }a_{n}\left( \omega \right) \int dK\, 
k^{\left\langle \mu\right. }k^{\left. \nu \right\rangle }
k^{\left\langle \alpha \right.}
k^{\left. \beta \right\rangle }E_{\mathbf{k}}^{n}f_{0\mathbf{k}}  \notag \\
&=&2\Delta ^{\mu \nu \alpha \beta }\sum_{n=0}^{\infty
}I_{n+4,2}\;a_{n}\left( \omega \right) \,,
\end{eqnarray}%
where we used \cite{DeGroot} 
\begin{equation}
\int dK\,k^{\left\langle \mu \right. }k^{\left. \nu \right\rangle
}k^{\left\langle \alpha \right. }k^{\left. \beta \right\rangle }E_{\mathbf{k}%
}^{n}f_{0\mathbf{k}}=\frac{2}{5!!}\,\Delta ^{\mu \nu \alpha \beta }\int
dK\,E_{\mathbf{k}}^{n}f_{0\mathbf{k}}\left( m^{2}-E_{\mathbf{k}}^{2}\right)
^{2}\;,  \label{56}
\end{equation}%
and introduced the thermodynamic function
\begin{equation}
I_{nq}=\frac{1}{\left( 2q+1\right) !!}\int dKf_{0\mathbf{k}}E_{\mathbf{k}%
}^{n-2q}\left( m^{2}-E_{\mathbf{k}}^{2}\right) ^{q}.  \label{57}
\end{equation}

Thus, Eq.\ (\ref{LinBoltz}) can be cast into a more convenient form 
\begin{equation}
\tilde{\pi}^{\mu \nu }\left( \omega ,\mathbf{0}\right) =2\tilde{G}_{R}\left(
\omega ,\mathbf{0}\right) \tilde{\sigma}^{\mu \nu }\left( \omega ,\mathbf{0}%
\right) ,  \label{Wow}
\end{equation}%
where we introduced the retarded Green's function%
\begin{equation}
\tilde{G}_{R}\left( \omega ,\mathbf{0}\right) =\sum_{n=0}^{\infty
}\,I_{n+4,2}\,a_{n}\left( \omega \right) \text{.}
\end{equation}%
Since in fluid dynamics the thermodynamic force related to the shear stress
tensor is defined as $2\tilde{\sigma}^{\mu \nu }$, we kept the factor $2$ in
Eq.\ (\ref{Wow}). As shown in the previous section, Eqs.\ (\ref{very nice})
and (\ref{very nice3}), the shear relaxation time is determined by the first
pole, $\omega _{1}$, of $\tilde{G}_{R}\left( \omega ,\mathbf{0}\right) $, as%
\begin{equation}
\tau _{\pi }=\frac{1}{i\omega _{1}\left( \mathbf{0}\right) },
\end{equation}%
while the shear viscosity coefficient is determined by the retarded Green's
function $\tilde{G}_{R}\left( \omega ,\mathbf{0}\right) $ at the origin, 
\begin{equation}
\eta =\left. \tilde{G}_{R}\left( \omega ,\mathbf{0}\right) \right\vert
_{\omega =0}\;.
\end{equation}%
Thus, the problem of finding the linear transport coefficients has been
reduced to determining the analytic properties of $a_{n}\left( \omega
\right) $.

Equations such as Eq.\ (\ref{equation}) appear quite often in problems
involving the extraction of transport coefficients from the Boltzmann
equation. The way to solve this problem is to substitute the expansion (\ref%
{expansion}) into Eq.\ (\ref{equation}), multiply by $E_{\mathbf{k}%
}^{m}k^{\left\langle \mu \right. }k^{\left. \nu \right\rangle }$, and
integrate over $dK$. Then one obtains 
\begin{eqnarray}
\lefteqn{\sum_{n=0}^{\infty }a_{n}\left( \omega \right) \int dKE_{\mathbf{k}%
}^{m}k^{\left\langle \mu \right. }k^{\left. \nu \right\rangle }\left[
-i\omega -\hat{C}\left( K\right) \right] f_{0\mathbf{k}}E_{\mathbf{k}%
}^{n}k^{\left\langle \alpha \right. }k^{\left. \beta \right\rangle }} 
\hspace*{2cm}  \notag \\
&=&\beta _{0}\int dKE_{\mathbf{k}}^{m-1}k^{\left\langle \mu \right.
}k^{\left. \nu \right\rangle }k^{\left\langle \alpha \right. }k^{\left.
\beta \right\rangle }f_{0\mathbf{k}}\;,
\end{eqnarray}%
Using Eqs.\ (\ref{56}) and (\ref{57}), we can rewrite this in the form 
\begin{equation}
\sum_{n=0}^{\infty }\left( -i\omega \mathcal{D}^{mn}+\mathcal{A}^{mn}\right)
a_{n}\left( \omega \right) =\beta _{0}I_{m+3,2}\;,
\end{equation}%
where we defined the matrices 
\begin{eqnarray}
\mathcal{A}^{mn}\Delta ^{\mu \nu \alpha \beta } &=&-\frac{1}{2}\int dKE_{%
\mathbf{k}}^{m}k^{\left\langle \mu \right. }k^{\left. \nu \right\rangle }%
\hat{C}\left( K\right) f_{0\mathbf{k}}E_{\mathbf{k}}^{n}k^{\left\langle
\alpha \right. }k^{\left. \beta \right\rangle }, \\
\mathcal{D}^{mn} &=&\frac{1}{5!!}\int dKf_{0\mathbf{k}}E_{\mathbf{k}%
}^{m+n}\left( m^{2}-E_{\mathbf{k}}^{2}\right) ^{2}\,.
\end{eqnarray}

Thus, the formal solution for $a_{n}(\omega )$ is 
\begin{equation}
a_{m}\left( \omega \right) =\beta _{0}\sum_{n=0}^{\infty }\left[ \left(
-i\omega \mathcal{D}+\mathcal{A}\right) ^{-1}\right] ^{mn}I_{n+3,2}
\end{equation}%
and the expression for $\tilde{G}_{R}\left( \omega ,\mathbf{0}\right) $
becomes%
\begin{equation}
\tilde{G}_{R}\left( \omega ,\mathbf{0}\right) =\beta _{0}\sum_{m=0}^{\infty
}\sum_{n=0}^{\infty }I_{m+4,2}\left[ \left( -i\omega \mathcal{D}+\mathcal{A}%
\right) ^{-1}\right] ^{mn}I_{n+3,2}.
\end{equation}%
The poles of the function above can be obtained from the roots of the
determinant 
\begin{equation}
\det \left( -i\omega \mathcal{D}+\mathcal{A}\right) =0\,,
\end{equation}%
which is given by the product of the eigenvalues $\lambda _{n}$ of the
operator $\left( -i\omega \mathcal{D}+\mathcal{A}\right) $. Note that,
because $\mathcal{D}$ and $\mathcal{A}$ are real matrices, all the poles are
on the imaginary axis. Thus, the truncation of the equation of motion to a
relaxation-type form is possible, if the separation between the poles is
large enough. The shear viscosity coefficient is given by 
\begin{equation}
\eta \equiv \left. \tilde{G}_{R}\left( \omega ,\mathbf{0}\right) \right\vert
_{\omega =0}=\beta _{0}\sum_{m=0}^{\infty }\sum_{n=0}^{\infty }I_{m+4,2}(%
\mathcal{A}^{-1})^{mn}I_{n+3,2}\,.
\end{equation}

Thus, in order to find the relaxation times and viscosity coefficients from
the linearized Boltzmann equation one has to invert and compute eigenvalues
of infinite matrices. In practice, however, one never deals with infinite
matrices because the expansion (\ref{expansion}) is always truncated, see,
e.g., Chapman-Enskog theory \cite{DeGroot}.

Let us consider the simplest possible case and take only one term in the
expansion (\ref{expansion}). We remark that this corresponds to using the
Israel-Stewart 14-moment approximation in the moments method. Then, $\tilde{G%
}_{R}\left( \omega ,\mathbf{0}\right) $ has the following simple form%
\begin{equation}
\tilde{G}_{R}\left( \omega ,\mathbf{0}\right) =\frac{i\beta _{0}I_{32}}{%
\omega +i\mathcal{A}^{00}/I_{42}}\;.
\end{equation}%
where we used that $\mathcal{D}^{00}=I_{42}$. In this case, the retarded
Green's function has only one pole, $\omega _{0}=-i\mathcal{A}^{00}/I_{42}$.
The relaxation time is obtained as 
\begin{equation}
\tau _{\pi }=\frac{1}{i\omega _{0}}=\frac{I_{42}}{\mathcal{A}^{00}}\;.
\end{equation}%
On the other hand, the shear viscosity is given by $\eta =\left. \tilde{G}%
_{R}\left( \omega ,\mathbf{0}\right) \right\vert _{\omega =0}$ and becomes 
\begin{equation}
\eta =\beta _{0}\frac{I_{42}I_{32}}{\mathcal{A}^{00}}\;.
\end{equation}%
In the massless limit, for a gas of hard spheres, one determines $\mathcal{A}%
^{00}$ to be 
\begin{equation}
\mathcal{A}^{00}=\frac{3}{5}\,I_{42}\,n_{0}\,\sigma \,,
\end{equation}%
where $\sigma $ is the total cross section and $n_{0}$ is the particle
number density. Then, 
\begin{eqnarray}
\eta &=&\frac{4}{3\sigma \beta _{0}}\;, \\
\tau _{\pi } &=&\frac{5}{3n_{0}\sigma }\;,
\end{eqnarray}%
where we used that, in the massless limit, 
\begin{eqnarray}
I_{42} &=&4\frac{P_{0}}{\beta _{0}^{2}}\;, \\
I_{32} &=&\frac{4}{5}\frac{P_{0}}{\beta _{0}}\;,
\end{eqnarray}%
where $P_{0}$ is the thermodynamic pressure. Also, in this particular
example, one can show that the ratio $\eta /\tau _{\pi }$ is independent of
the cross section, 
\begin{equation}
\frac{\eta }{\tau _{\pi }}=\beta _{0}I_{32}\;.
\end{equation}%
These are exactly the results obtained in Ref.\ \cite{dkr}. This
demonstrates that the relaxation time in IS theories, which determines the
time scale of the transient dynamics of the dissipative currents, is indeed
a microscopic and not a fluid-dynamical time scale. It is determined by the
interparticle scattering rate, and not by the time scales of fluid dynamics
located near the origin of the complex $\omega-$plane. We shall demonstrate
in Sec.\ \ref{Discussion} that attempts to extract the value of the
relaxation time from the dynamics on fluid-dynamical time scales in general
fail to give the correct expression.

It is important to remark that by including more terms in the expansion (\ref%
{expansion}) we obtain a retarded Green's function with more poles.
Furthermore, the expression for the first pole and, consequently, the
relaxation time, will also be modified. All the other transport coefficients
will also receive corrections.

\subsection{Linear Response Theory and Metric Perturbations}

\label{metricsection}

We now apply our formalism to the case studied in Ref.\ \cite{BRSSS}, where
the transport coefficients are determined from perturbations $h^{\mu \nu}$
of the metric tensor 
\begin{equation}
g^{\mu \nu }=\eta ^{\mu \nu }+h^{\mu \nu }\text{.}  \label{metric_pert}
\end{equation}%
This method can be equally applied at strong and weak coupling. The
variation of the energy-momentum tensor $T^{\mu \nu}$ due to the metric
perturbations is \cite{starinets} 
\begin{equation}
\delta T^{\mu \nu } \left( X\right) =\frac{1}{2}\int_{-\infty }^{\infty
}d^{4}X^{\prime }\,G_{R}^{\mu \nu \alpha \beta }\left( X-X^{\prime }\right)
\,h_{\alpha \beta }\left( X^{\prime }\right) ,
\end{equation}%
where $G_{R}^{\mu \nu \alpha \beta }\left( X-X^{\prime}\right) $ is the
retarded Green's function.

For the sake of simplicity, a very peculiar type of metric perturbation is
considered, $h_{xy}=h_{xy}\left( t,z\right) $, with all other components of
the metric tensor left unperturbed \cite{BRSSS,guymoore}. For this specific
type of metric perturbation, all the other components of $\delta T^{\mu \nu
} $ decouple from the $xy$ component and one arrives at a very simple
expression for $\delta T^{xy} $ \cite{starinets} 
\begin{equation}
\delta T^{xy} (t,z) = \int_{-\infty }^{\infty }dt^{\prime }\,dz^{\prime }
\,G_{R}^{xyxy}\left(t-t^{\prime };\,z-z^{\prime }\right) h_{xy}\left(
t^{\prime },z^{\prime }\right) .
\end{equation}

The energy-momentum tensor $T^{\mu \nu }$ is then assumed to have the
traditional fluid-dynamical structure 
\begin{equation}
T^{\mu \nu }=\varepsilon \,u^{\mu }u^{\nu }-\Delta ^{\mu \nu }P+\pi ^{\mu
\nu }\;.
\end{equation}%
For the particular metric perturbation considered here, the effects of bulk
viscosity can be neglected without loss of generality. It is possible to
show that 
\begin{equation}
\delta T^{xy}\equiv T^{xy}(\eta ^{\mu \nu }+h^{\mu \nu })-T^{xy}(\eta ^{\mu
\nu })=-P_{0}\,h^{xy}+\delta \pi ^{xy}\;.
\end{equation}%
where $\delta \pi ^{xy}$ is the $xy$ component of the shear stress tensor
generated by the metric perturbations and $P_{0}$ is the pressure of the
unperturbed state. For the sake of simplicity, the shear stress tensor of
the unperturbed state is set to zero. Also, it is easy to see, using the
equations of motion, that the velocity terms in the energy-momentum tensor
do not contribute to $\delta T^{xy}$. Thus, we arrive at the following
equation 
\begin{equation}
\delta \pi ^{xy}=P_{0}\,h^{xy}+\int_{-\infty }^{\infty }dt^{\prime
}\,dz^{\prime }\,G_{R}^{xyxy}\left( t-t^{\prime };\,z-z^{\prime }\right)
h_{xy}\left( t^{\prime },z^{\prime }\right) .
\end{equation}%
After taking the Fourier transform we can show that 
\begin{equation}
\delta \tilde{\pi}^{xy}(Q)=\tilde{G}_{R}(Q)\,\tilde{h}_{xy}(Q)\;,
\label{linearresponsemetric}
\end{equation}%
where $\tilde{G}_{R}(Q)=-P_{0}+\tilde{G}_{R}^{xyxy}(Q)$. Note that $%
h^{xy}=-h_{xy}$. Since the pressure has no dependence on $Q$, it is clear
that $\tilde{G}_{R}(Q)$ has the same analytic structure as $\tilde{G}%
_{R}^{xyxy}(Q)$.

Assuming that $\tilde{G}_{R}^{xyxy}$ has $N$ poles, $\omega _{i}(\mathbf{q})$%
, and that the first pole is purely imaginary, we can apply Eq.\ (\ref{38})
and obtain the equation of motion for $\delta \pi ^{xy}$, 
\begin{equation}
\tau _{\pi }\partial _{t}\delta \pi ^{xy}+\delta \pi
^{xy}=D_{0}h_{xy}+D_{1}\partial _{t}h_{xy}+D_{2}\partial _{t}^{2}h_{xy}+%
\mathcal{O}\left( \partial _{t}^{3}h_{xy},\partial _{z}^{2}h_{xy}\right) \;.
\label{ISmetric}
\end{equation}%
Note that if the first pole is not purely imaginary, a simple relaxation
equation would not be able to describe the transient dynamics, even for long
times. In Eq.\ (\ref{ISmetric}), we introduced the following transport
coefficients, 
\begin{eqnarray}
\tau _{\pi } &=&\frac{1}{i\omega _{1}\left( \textbf{0}\right) }\;,  \notag \\
D_{0} &=&\left. \tilde{G}_{R}\left( \omega ,\textbf{0}\right) \right\vert _{\omega
=0}=-P_{0}+\left. \tilde{G}_{R}^{xyxy}(\omega ,\textbf{0})\right\vert _{\omega
=0}\equiv -P_{0}+P_{0}=0\;,  \notag \\
D_{1} &=&i\partial _{\omega }\left. \tilde{G}_{R}\left( \omega ,\textbf{0}\right)
\right\vert _{\omega =0}+\tau _{\pi }\,D_{0}=i\partial _{\omega }\left. 
\tilde{G}_{R}\left( \omega ,\textbf{0}\right) \right\vert _{\omega =0}\equiv \eta 
\text{ },  \notag \\
D_{2} &=&-\frac{1}{2}\partial _{\omega }^{2}\left. \tilde{G}_{R}\left(
\omega ,\textbf{0}\right) \right\vert _{\omega =0}+D_{1}\tau _{\pi }-D_{0}\tau _{\pi
}^{2}\equiv -\frac{1}{2}\partial _{\omega }^{2}\left. \tilde{G}_{R}\left(
\omega ,\mathbf{0}\right) \right\vert _{\omega =0}+\eta \tau _{\pi }\;,
\label{matchingconditionsISlinear}
\end{eqnarray}%
The expression for the shear viscosity $\eta $ is now different from the one
in the Boltzmann case, because $2\sigma ^{xy}\equiv \partial _{t}h_{xy}$, so
that the coefficient of the shear tensor is actually $D_{1}$ (and not $D_{0}$%
, as before). On the other hand, the relaxation time is still given by the
inverse of the first pole.

\section{Discussion}

\label{Discussion}

Relaxation-type equations for the dissipative currents have been recently
derived in Ref.\ \cite{BRSSS}. We give a brief account of the strategy
employed in that work in terms of our notation. The starting point is the
gradient expansion (\ref{ExpJ}), assuming that a Knudsen number counting as
explained at the end of Sec.\ \ref{gradientsection} is applicable. The
gradient expansion (\ref{ExpJ}) is not an equation of motion for the
dissipative current, but one can construct one by taking the first-order
solution, $J = \bar{D}_0\, F + \mathcal{O}(\mathrm{Kn}^2)$, and then
replacing the first time derivative of $F$ on the right-hand side by a time
derivative of $J$, 
\begin{equation}
J = \bar{D}_0 \, F + \bar{D}_1 \, \partial_t \left( \frac{J}{\bar{D}_0}
\right) + \mathcal{O}(\mathrm{Kn}^3) \;\;\;\; \Longleftrightarrow \;\;\;\; 
\bar{\tau}_R\, \partial_t J + J \simeq \bar{D}_0\, F\;,
\end{equation}
where $\bar{\tau}_R \equiv -\bar{D}_1/\bar{D}_0$ has the dimension of time.
By construction, this is a relaxation-type equation of motion for the
dissipative current $J$, with a relaxation time $\bar{\tau}_R$.

From our previous discussion, however, it is clear that this need not be the
correct equation of motion for the dissipative current. If the poles of the
retarded Green's function $\tilde{G}_R(\omega)$ associated with the
dissipative current $J$ are off the imaginary axis, the equation of motion
for $J$ is never of relaxation type, and the above way to construct one is
misleading, since it fails to capture the correct physics. Only if the first
pole of $\tilde{G}_R(\omega)$ lies on the imaginary axis and is sufficiently
separated from the other singularities, one can obtain a relaxation-type
equation for $J$. In this case, however, the true relaxation time is $\tau_R
= 1/[i\omega_1(\mathbf{0})]$, and not $\bar{\tau}_R = -\bar{D}_1/\bar{D}_0$.

There is, however, a particular case, where $\tau_R = \bar{\tau}_R$, namely
when $D_1 =0$, i.e., when $\Xi(\omega, \mathbf{0}) = const.$. This is most
easily seen in the one-pole case, where, cf.\ Eq.\ (\ref{CoeffsA}), 
\begin{equation}
0 = i \left. \partial_\omega \tilde{G}_R(\omega,\mathbf{0})
\right|_{\omega=0} + D_0\, \tau_R \equiv \bar{D}_1 + \bar{D}_0\, \tau_R\;,
\end{equation}
i.e., $\tau_R = -\bar{D}_1/\bar{D}_0 \equiv \bar{\tau}_R$.

This argument can also be applied to the case discussed in Sec.\ \ref%
{metricsection}, where the thermodynamic force is not given by $F$ but by $%
\partial_t F$. Then, $D_0=0$, and $D_1 = \bar{D}_1 = \eta$ is the transport
coefficient. In this case, $\bar{\tau}_\pi = - \bar{D}_2/\bar{D}_1$, and
equivalency to the true relaxation time requires that $D_2 = 0$, i.e., 
\begin{equation}
0 = -\frac{1}{2}\partial _{\omega }^{2}\left. \tilde{G}_{R}\left( \omega ,%
\mathbf{0}\right) \right\vert _{\omega =0} + D_1\, \tau_\pi \equiv \bar{D}_2
+ \bar{D}_1\, \tau_\pi\;,
\end{equation}
or 
\begin{equation}
\eta\, \tau_\pi = \frac{1}{2}\partial _{\omega }^{2}\left. \tilde{G}%
_{R}\left( \omega ,\mathbf{0}\right) \right\vert _{\omega =0}\;.
\end{equation}
This equation is rather similar to the one given in Refs.\ \cite%
{BRSSS,guymoore}, $\eta \, \bar{\tau}_\pi = [ \partial _{\omega }^{2}\tilde{G%
}_{R}(0)-$ $\partial _{q_{z}}^{2}\tilde{G}_{R}(0)]/2 $ \cite{comment2}. Here, the additional
derivatives with respect to momentum enter because second-order time
derivatives also arise from space-time curvature in Eq.\ (\ref{ExpJ}) (they
were not explicitly denoted in that equation), cf.\ Ref.\ \cite{BRSSS}.
These are not subjected to the construction of a relaxation-type equation
for $J$ as explained above. In turn, $\eta \bar{\tau}_\pi$ receives an
additional contribution from second-order spatial derivatives, for details,
see Ref.\ \cite{BRSSS}.

Regardless of these considerations, the true relaxation time is always given
by the first pole of the retarded Green's function. In general, the location
of this pole cannot be found from a \textit{truncated\/} Taylor expansion
around the origin.

A calculation of the shear viscosity relaxation time coefficient has also
been performed in Refs.\ \cite{tomoi} within the Mori-Zwanzig formalism \cite%
{mori}. The implications of our findings to those works will be discussed in
detail elsewhere.

\section{Summary}

\label{conclusion}

In this work, we have derived equations of motion for the dissipative
currents, assuming these currents to be linearly related to the
thermodynamic forces. We have shown how these equations of motion are
determined by the analytical structure of the associated retarded Green's
function in the complex $\omega$ plane. We have demonstrated that the
standard gradient expansion is equivalent to a Taylor expansion of the
retarded Green's function around the origin in the complex $\omega$ plane.
This Taylor series is convergent only when all singularities of the retarded
Green's function are pushed to infinity. In general, however, these
singularities appear at finite values of $|\omega|$, which consequently
severely restricts the applicability of the gradient expansion.

We have furthermore demonstrated that, if the retarded Green's function has
simple poles in the complex $\omega$ plane, the dissipative current obeys a
differential equation with source terms which are the thermodynamic force
and gradients thereof. This is different from the gradient expansion where
the current is directly proportional to the thermodynamic force and its
gradients. In general, the equation of motion for the dissipative current is
not a relaxation-type equation. However, in the limit where all
singularities of the retarded Green's function except the pole nearest to
the origin are pushed to infinity, it is possible to approximate the
dynamical equation satisfied by $J$ as a relaxation-type equation, similar
to the ones appearing in Israel-Stewart theory. This is only possible if the
first pole is purely imaginary. The relaxation time is equal to minus the
inverse of the imaginary part of the pole. The gradient expansion
constitutes the asymptotic solution of the resulting relaxation-type
equation, and can be obtained by taking the relaxation time to zero or,
equivalently, pushing the first pole to infinity. This is consistent with
the above statement that the gradient expansion arises from a Taylor
expansion of the retarded Green's function around the origin.

In relativistic systems, in order to have stable and causal equations of
motion for the dissipative currents one cannot take the relaxation time to
be arbitrarily small \cite{us}. Thus, one cannot push the first pole of the
retarded Green's function to infinity. This is the reason why one cannot use
the gradient expansion to obtain stable and causal equations of motion for
the dissipative currents.

As an example, we have studied the Boltzmann equation for a classical gas
and demonstrated that the retarded Green's function associated with the
shear stress tensor has infinitely many simple poles along the imaginary
axis. Therefore, it is possible to reduce the equation of motion for the
shear stress tensor to a relaxation-type equation. Our results are
consistent with those of Ref.\ \cite{dkr}, when one truncates the collision
operator at lowest order. This convincingly demonstrates that the
time scale of transient dynamics determined by the relaxation time is of
microscopic, and not of fluid-dynamical origin: it is the slowest
microscopic time scale, not the fastest fluid-dynamical time scale as
implicitly assumed in attempts to extract the relaxation time by expanding
the retarded Green's function around the origin. Consequently, the
expression for the relaxation time derived in Refs.\ \cite{BRSSS,guymoore}
by using an expansion around the origin is, in general, different from the
one derived from the first pole of the retarded Green's function. In fact,
as we have demonstrated in this work, when the retarded Green's function has
simple poles off the imaginary axis in the complex $\omega$ plane, the true
dynamics of the system at long wavelengths and low frequencies is not even
of relaxation type. Strongly coupled theories, like those emerging from the
AdS/CFT correspondence \cite{starinets}, belong to this class.

\section*{Acknowledgement}

The authors thank T.~Kodama, T.~Koide, A.~Ficnar, P.~Romatschke, and
G.~Torrieri for fruitful discussions. J.N.\ acknowledges support from DOE
under Grant No.\ DE-FG02-93ER40764 and FAPERJ. H.N.\ was supported by the
ExtreMe Matter Institute EMMI. The authors thank the Helmholtz International
Center for FAIR within the framework of the LOEWE program for support.


\begin{thebibliography}{99}
\bibitem{comment1} Even though we only show terms up to order $\mathcal{O}%
\left( \mathrm{Kn}^{3}\right) $ in Eq.\ (\ref{Foda}), this equivalence can
be easily confirmed to all orders.
\bibitem{comment2} Note that this result was derived using a different metric
signature, i.e., $\eta _{\mu \nu }=\mathrm{diag}\,(-,+,+,+)$. See Ref.\ \cite{BRSSS} for details.
\end{thebibliography}
\end{document}